\newcommand{\CLASSINPUTtoptextmargin}{0.75in}
\newcommand{\CLASSINPUTbottomtextmargin}{1in}
\newcommand{\CLASSINPUTinnersidemargin}{0.625in}
\newcommand{\CLASSINPUToutersidemargin}{0.625in}
\documentclass[10pt, conference]{IEEEtran}
\IEEEoverridecommandlockouts
\usepackage{cite}
\usepackage{amsmath,amssymb,amsfonts}
\usepackage{algorithmic}
\usepackage{graphicx}
\usepackage{textcomp}
\usepackage{xcolor}
\usepackage{soul}
\usepackage{graphicx}
\usepackage[ruled,vlined]{algorithm2e}
\usepackage{subcaption}
\usepackage{listings}
\usepackage{makecell}
\usepackage{hyperref}

\def\BibTeX{{\rm B\kern-.05em{\sc i\kern-.025em b}\kern-.08em
    T\kern-.1667em\lower.7ex\hbox{E}\kern-.125emX}}
\SetAlFnt{\scriptsize}

\begin{document}

\title{Accelerating Trust Convergence in IIoT: A ML Approach for Dynamic Network Conditions\thanks{\textcopyright\ 2025 IEEE. Published in IEEE GLOBECOM 2025, pp.\ 4427--4432. DOI: 10.1109/GLOBECOM59602.2025.11431874}}

\author{
	\IEEEauthorblockN{Aymen Bouferroum}
	\IEEEauthorblockA{
		\textit{Inria Lille-Nord Europe}\\
		Lille, France\\
		aymen-salah-eddine.bouferroum@inria.fr
	}
	
	\and
	\IEEEauthorblockN{Valeria Loscri}
	\IEEEauthorblockA{
		\textit{Inria Lille-Nord Europe}\\
		Lille, France\\
		valeria.loscri@inria.fr
	}
	
	\and
	\IEEEauthorblockN{Abderrahim Benslimane}
	\IEEEauthorblockA{
		\textit{LIA/CERI University of Avignon}\\
		Avignon, France\\
		abderrahim.benslimane@univ-avignon.fr
	}
}

\maketitle

% --------------------------------------- Abstract -----------------------------------------------
\begin{abstract}
In Industrial Internet of Things (IIoT) environments, trust management plays a vital role in securing systems, especially when dealing with resource-constrained devices. Traditional trust models often overlook the impact of fluctuating network quality, leading to slower trust convergence and inaccurate assessments. In this paper, we propose a dynamic trust management solution, known as the Trust Convergence Acceleration (TCA) approach, which integrates Machine Learning (ML) to accelerate trust convergence under poor network conditions. Our model predicts the number of time units needed for trust convergence based on key network metrics and dynamically adapts transition probabilities in the trust model to enhance convergence speed. Using a simulation framework that incorporates realistic Wi-Fi channel conditions based on the IEEE 802.11 standard, we demonstrate the effectiveness of the TCA-based approach, achieving up to a 28.6\% reduction in trust convergence time under challenging conditions. Furthermore, the proposed solution exhibits resilience in scenarios involving malicious nodes, improving trust evaluation accuracy. This work provides a scalable and adaptive trust framework for IIoT systems in dynamic industrial environments, ensuring robust performance under varying network conditions.
\end{abstract}


% --------------------------------------- Section I -----------------------------------------------
\section{Introduction}

The rapid advancement of the Industrial Internet of Things (IIoT) has transformed industrial sectors through Industry 4.0 \cite{sadeghi2015security}, integrating cyber-physical systems, cloud computing, and artificial intelligence into manufacturing processes to enhance operational efficiency and decision-making capabilities \cite{boudagdigue2020trust}. This evolution, however, introduces significant security challenges as IIoT networks grow increasingly complex, particularly in resource-constrained environments where traditional encryption-based security proves impractical \cite{serror2021challenges}. The heterogeneous nature of these devices, with their limited computational and energy resources, has led to the emergence of trust management systems as a viable alternative for securing industrial operations. These systems \cite{serror2021challenges} continuously monitor device behavior, assigning and updating trust levels based on interactions and performance \cite{boudagdigue2018distributed}, enabling effective detection and isolation of potentially compromised devices.

Trust management approaches can be categorized into centralized, distributed, and hybrid models \cite{boudagdigue2020trust, alshehri2018clustering, chen2015trust}. Centralized models provide global network visibility but face scalability challenges. Distributed models offer improved scalability but require significant energy consumption and time to achieve trust convergence \cite{abderrahim2017tmcoi}, which is problematic in resource-constrained environments. Hybrid models, such as the H-IIoT architecture \cite{boudagdigue2020trust}, combine the strengths of both approaches by organizing IIoT devices into industrial communities managed by trusted leaders, balancing localized trust management with a global perspective.

In operational IIoT environments, network quality is subject to continuous variations caused by electromagnetic interference, device mobility, and network congestion. These fluctuations significantly impact trust management systems, as network metrics such as availability, throughput, and delay influence the effectiveness of trust evaluations \cite{Mohammadi2019}. While some researchers have employed recommendation filtering algorithms to address varying contextual factors \cite{CHEN2021107952} or introduced fixed probability models for network constraints \cite{boudagdigue2020trust}, these approaches often fail to capture the nuanced impact of dynamic network quality on trust evaluation. Consequently, trust management systems lacking real-time adaptation risk delayed detection of malicious activities and inaccurate assessments of device trustworthiness.

To address these challenges, we propose the Trust Convergence Acceleration (TCA), a novel trust management solution designed to enhance trust metric convergence by predicting and optimizing the trust evaluation process under varying network conditions. Our TCA integrates the Random Forest algorithm, which offers optimal performance with limited data, making it particularly suitable for resource-constrained IIoT environments. Empirical evaluations confirm its superior performance in terms of prediction accuracy and convergence speed. Integrated within the H-IIoT architecture \cite{boudagdigue2020trust} and leveraging Wi-Fi 6 (IEEE 802.11ax) technology \cite{wba2022wifi}, \cite{IEEE80211ax}, our framework provides an adaptive, scalable, and robust solution that significantly improves performance under fluctuating network conditions. Our work makes the following key contributions:

\begin{itemize}
    \item Development of a trust management model that integrates real-time network channel conditions into trust evaluations, capturing the inherent variability of IIoT environments.
    \item Design and implementation of a Machine Learning (ML) framework to accelerate trust convergence under varying network conditions, ensuring rapid and reliable system performance.
    \item Demonstration of enhanced robustness against malicious nodes, showcasing the resilience of our approach against adversarial behaviors.
\end{itemize}

The rest of this paper is organized as follows: Section II introduces the system model and problem formulation. Section III details our ML-enhanced trust management solution. Section IV evaluates the performance of our approach. Finally, Section V concludes and outlines future research directions.
% ----------------------------------- Section II -------------------------------------------

\section{System Model and Problem Formulation}

Our proposed trust management approach requires a formal characterization of the IIoT environment to address the challenges of dynamic industrial networks. We establish a comprehensive framework encompassing network architecture, trust formulation mechanisms, and an analysis of how varying network quality influences trust convergence in operational settings.

%                 -------------------------- Subsection A ----------------------------

\subsection{Network Architecture}
Our system model builds upon an enhanced version of the Hybrid Industrial Internet of Things (H-IIoT) architecture \cite{boudagdigue2020trust}, integrating Wi-Fi 6 technology to meet the demands of modern industrial environments. The architecture organizes IIoT devices into managed industrial communities, providing a foundation for efficient trust management in large-scale deployments.
\vspace{-10pt}
\begin{figure}[ht]
\centering
\includegraphics[ scale = 0.17]{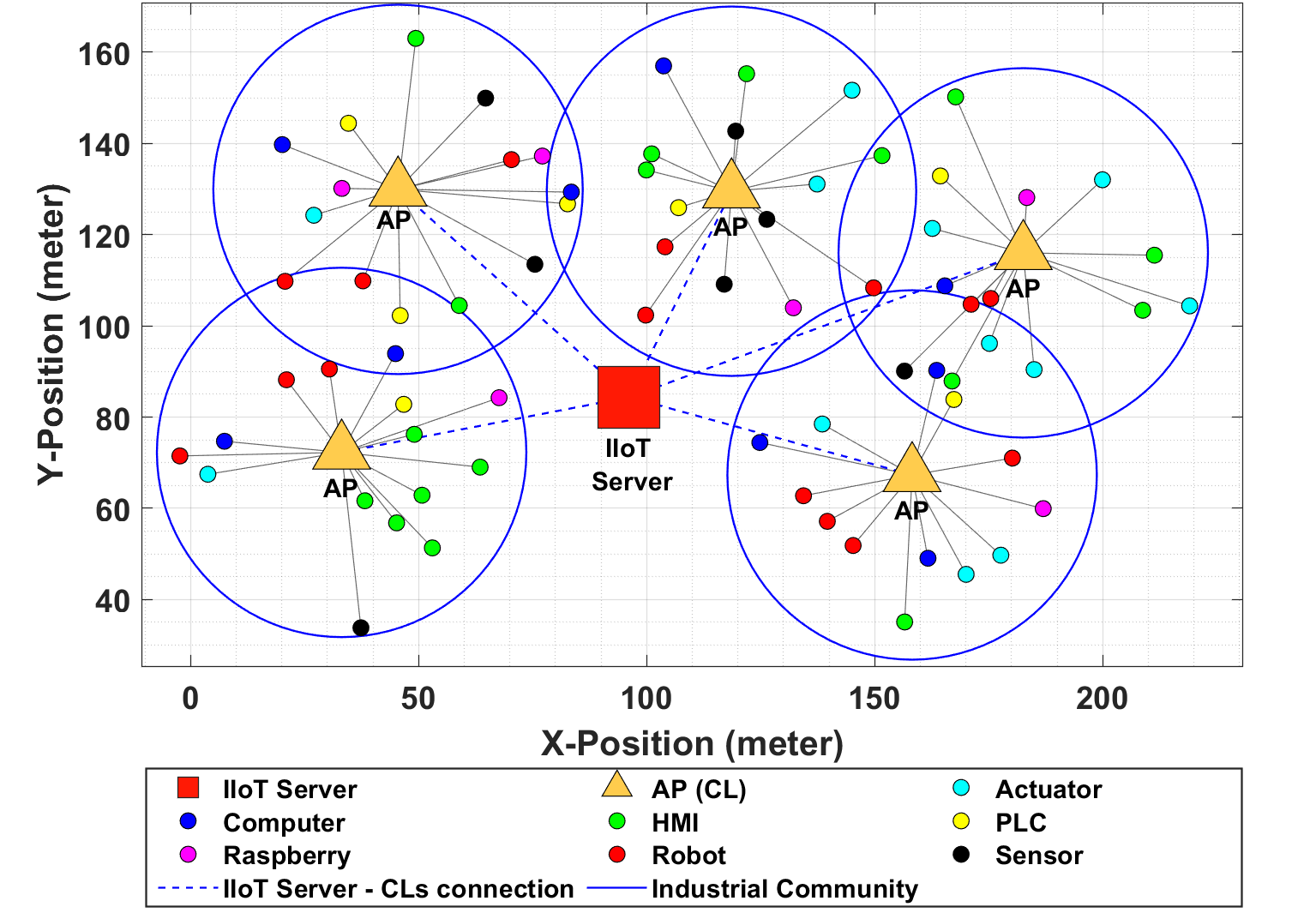}
% \includegraphics[clip, trim= 40 40 65 135, scale = 0.36]{figures/architecture_figure_new.eps}
% %
\caption{Network architecture.}
\label{fig:network_architecture}
\vspace{-10pt}
\end{figure}

The network consists of \(K\) Community Leaders (CLs) and \(L\) Member Nodes (MNs), denoted by \(\text{CL}_j | j \in \{1, 2, \ldots, K\}\) and \(\text{MN}_i | i \in \{1, 2, \ldots, L\}\), respectively. The main components include:
\begin{enumerate}
\item \textbf{IIoT Server}: A cloud-based central authority that aggregates global trust metrics and ensures consistent trust evaluation across the network.
\item \textbf{CLs}: Wi-Fi 6 Access Points (APs) serving as trusted supervisors for industrial communities, each managing a Basic Service Set (BSS) of devices.

\item \textbf{MNs}: Various IIoT devices (sensors, robots, actuators, PLCs, HMIs) organized into communities based on proximity and operational roles.

\item \textbf{Industrial Communities}: Functional groupings of MNs managed by a single CL, corresponding to a BSS in Wi-Fi terminology, enabling localized trust management and reduced communication overhead.
\end{enumerate}

For simulation purposes, our environment comprises a total of 200 nodes distributed across a 200 x 200 m² area representing an industrial facility. From these, 10 nodes function as CLs, while the remaining 190 operate as MNs with varying industrial roles. This scale was selected to reflect realistic industrial deployments while remaining computationally feasible for detailed network quality analysis. Fig. \ref{fig:network_architecture} provides a simplified visualization of this architecture.
The hierarchical structure created by these components enables efficient trust evaluation through a balance of localized management and global coordination. CLs monitor their associated nodes while communicating with the IIoT server to maintain network-wide trust consistency, significantly reducing overhead in large-scale deployments.
Our implementation leverages IEEE 802.11 standards due to their widespread adoption in industrial settings, technical maturity, and robust performance in challenging electromagnetic environments. Specifically, we selected Wi-Fi 6 for its advanced capabilities that address critical IIoT requirements \cite{wba2022wifi}. These include Orthogonal Frequency Division Multiple Access (OFDMA) for improved multi-device communication, Multi-User Multiple Input Multiple Output (MU-MIMO) for enhanced throughput, BSS coloring to reduce interference in dense deployments, and Target Wake Time (TWT) for energy efficiency. While our current implementation utilizes Wi-Fi 6, the framework's modular design ensures compatibility with emerging wireless technologies. Note that the centralized server, while enabling global coordination, represents a potential single point of failure that can be mitigated through redundancy or edge-based failover in critical deployments.
%                 -------------------------- Subsection B ----------------------------
\subsection{Trust Model Formulation}
\label{subsec:trust_model}

The trust model, Tm-IIoT, evaluates the trustworthiness of IIoT devices based on their behavior and interactions within the network \cite{boudagdigue2020trust}. Each node $i$ is assigned a trust metric $Tm_i$, continuously updated based on these performance metrics:

\begin{enumerate}
    \item \textbf{Cooperation Rate (\( {C_{MN_i}} \))}: Reflects the node's willingness to cooperate in network activities (e.g., forwarding messages):
        \begin{equation}
            \scalebox{0.9}{$
            \begin{gathered}
                C_{MN_i} = \frac{\sum_{i=1}^{NF} C_{mi}}{NF},
            \end{gathered}$}
        \end{equation}
    where \(NF\) is the total number of messages transmitted to the node, and \(C_{mi}\) is 1 if message \(mi\) is forwarded correctly, 0 otherwise.

    \item \textbf{Direct Honesty (\( {D_{MN_i}} \))}: Assesses how well the node's activities align with its expected operational behavior, as defined in its device configuration profile:
        \begin{equation}
            \scalebox{0.9}{$
            \begin{gathered}
                D_{MN_i} = \frac{|A(MN_i) \cap AF(MN_i)|}{|A(MN_i) \cup AF(MN_i)|},
            \end{gathered}$}
        \end{equation}
    where \(A(MN_i)\) is the set of activities the node should perform according to its operational role, and \(AF(MN_i)\) is the set of activities actually performed.

    \item \textbf{Indirect Honesty (\( {I_{MN_i}} \))}: Reflects the node's reputation within its community, based on feedback from other nodes:
        \begin{equation}
            \scalebox{0.9}{$
            \begin{gathered}
                I_{MN_i} = \frac{\sum_{MN_i'=1}^{N} R_{MN_i',MN_i}}{N},
            \end{gathered}$}
        \end{equation}
    where \(R_{MN_i',MN_i}\) is the reputation given by node \(MN_i'\) to node \(MN_i\), and \(N\) is the total number of MNs. This peer-evaluation mechanism helps identify malicious nodes that might appear legitimate in direct evaluations.
\end{enumerate}

The model employs a finite-state discrete-time Markov chain with $(E+1)$ states to represent trust evolution, where $E$ is set to 10 in our implementation. Each state corresponds to a discrete trust value between 0 (completely untrusted) and 1 (fully trusted), with uniform increments of $\phi = 0.1$. 

While continuous trust models offer more granularity, we adopt a discrete approach for three key reasons: (1) computational efficiency in resource-constrained environments, (2) increased robustness against small fluctuations in trust metrics, and (3) simplified decision-making for trust-based actions. The choice of $E{=}10$ states balances granularity with stability: coarser ($E{=}5$) and finer ($E{=}20$) discretizations in our tests preserved qualitative outcomes while $E{=}10$ minimized oscillations.

The transition matrix $P$ of the Markov chain is defined as:
\vspace{-3pt}
\begin{equation}
    \scalebox{0.9}{$
    \begin{gathered}
        P = (P_{i,j}(t))_{0 \leq i,j \leq E},
    \end{gathered}$}
\end{equation}
where $P_{i,j}(t)$ represents the probability of transitioning from state $i$ to state $j$ at time $t$, expressed as:
\vspace{-3pt}
\begin{equation}
    \scalebox{0.9}{$
    \begin{gathered}
       P_{i,j}(t) = Pr(Y_t = j | Y_{t-1} = i),
    \end{gathered}$}
\end{equation}
where $Y_t$ is the random variable representing the trust metric $Tm$ of an IIoT device at time $t$. The probability of being in state $i$ at time $t$, given an initial state $Y_0 = 1$, is:
\vspace{-3pt}
\begin{equation}
    \scalebox{0.9}{$
    \begin{gathered}
        p_i(t) = \sum_{z \in [1...E]} P_{1,z}(t_z) \cdot P_{z,i}(t).
    \end{gathered}$}
\end{equation}

At each time step, the trust calculation allows for five possible transitions based on the performance metrics: increasing, decreasing, remaining in the same state, reaching the maximum trust state ($E$), or falling to the untrusted state ($0$). This mechanism ensures adaptive trust assessment, enabling the model to respond to changes in node behavior over time.

% \begin{figure}[ht]
%     \centering
%     \includegraphics[width=1\linewidth]{figures/State_transition.pdf}
%     \caption{State transition diagram for Tm-IIoT model.}
%     \label{fig:state_transition}
%     \vspace{-15pt} 
% \end{figure}

%                 -------------------------- Subsection C ----------------------------
\subsection{Network Condition Modeling and $Tm$ Convergence Analysis}
\label{subsec:secIIC}

The assessment of trust in Wi-Fi 6 environments necessitates consideration of Quality of Service (QoS) parameters identified in \cite{Sugeng2015THEIO} as significant for network performance and trust evaluation. These parameters include: Signal-to-Noise Ratio (SNR), Packet Loss probability (PL), Jitter (J), Latency (L), Throughput (T), and Signal-to-Interference-plus-Noise Ratio (SINR). Each parameter directly influences data transmission reliability, which subsequently affects trust evaluation accuracy. Our investigation leverages MATLAB's WLAN and Communications Toolboxes for creating this high-fidelity simulation environment, selected for their accurate implementation of IEEE 802.11ax specifications. This framework incorporates detailed physical layer characteristics, including OFDMA resource allocation and MU-MIMO transmission schemes, ensuring that network behavior faithfully represents operational conditions in industrial settings.

We conducted simulations analyzing three distinct network condition scenarios (Good, Medium, and Poor) derived from IEEE 802.11ax standard specifications and observed performance characteristics in industrial Wi-Fi 6 deployments. Table \ref{tab:simulation_scenarios} summarizes these parameter ranges, which represent realistic operational conditions in industrial environments based on field measurements and standard performance metrics for Wi-Fi 6 networks.

\begin{table}[ht]
\vspace{-4pt}
    \caption{Network condition scenarios with parameter ranges.}
    \label{tab:simulation_scenarios}
    \centering
    \resizebox{0.95\linewidth}{!}{%
        \begin{tabular}{|c|c|c|c|}
            \hline
            \textbf{Parameter} & \textbf{Scenario 1} & \textbf{Scenario 2} & \textbf{Scenario 3} \\
            & \textbf{Good} & \textbf{Medium} & \textbf{Poor} \\
            \hline
            SNR (dB) & 30--41 & 20--29 & 10--19 \\
            \hline
            Packet Loss Probability (\%) & 0--1 & 1--4 & 4--8 \\
            \hline
            Jitter (ms) & 0--2 & 2--10 & 10--30 \\
            \hline
            Latency (ms) & 1--5 & 5--20 & 20--50 \\
            \hline
            Throughput (\% of max) & 80--100 & 50--80 & $<$50 \\
            \hline
            SINR (dB) & $>$25 & 15--25 & $<$15 \\
            \hline
        \end{tabular}%
    }
    \vspace{-10pt}
\end{table}

Our simulation methodology for trust convergence analysis comprised the following key steps:
\begin{enumerate}
    \item \textbf{Baseline Establishment}: We first conducted baseline simulations under ideal network conditions to determine the stable trust value for each node in the network, representing its "ground truth" trustworthiness.
    \item \textbf{Trust Perturbation}: We then artificially set the trust values of selected nodes to 0.5, deliberately deviating from their established ground truth values.
    \item \textbf{Convergence Monitoring}: Under each network condition scenario, we measured the number of time units required for the perturbed trust values to recover and stabilize at their ground truth values.
\end{enumerate}

\begin{figure}[ht]
    \vspace{-5pt}
    \centering
    \includegraphics[width=0.59\linewidth]{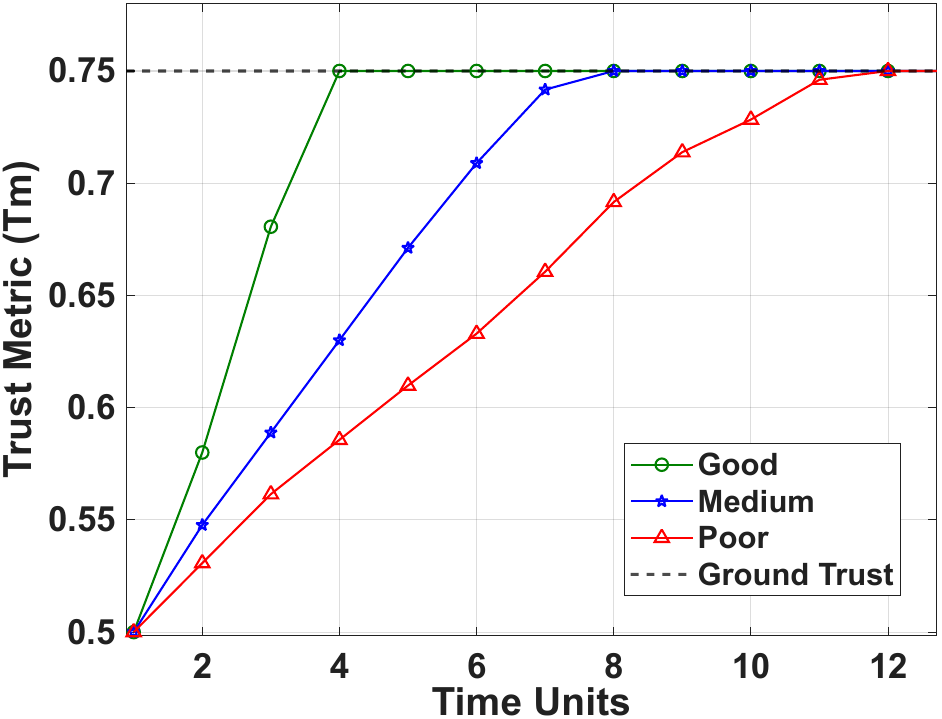}
    \caption{$Tm$ convergence over time under varying network conditions.}
    \label{fig:convergence_speed}
    \vspace{-5pt}
\end{figure}

Fig. \ref{fig:convergence_speed} illustrates trust evolution across the three scenarios, revealing a clear correlation between network quality and trust stabilization time. Under Good conditions, trust recovers from perturbation ($Tm$=0.5) to ground truth ($Tm$=0.75) in only 4 time units, compared to 8 units in Medium conditions and 12 in Poor conditions. When trust converges under Good conditions, under Poor conditions it has only reached approximately 0.58, creating periods of uncertainty about node trustworthiness that compromise security.
These findings underscore the need for an adaptive trust management approach that adapts to network conditions to achieve faster trust convergence without compromising evaluation accuracy.
% --------------------------------------- Section III ----------------------------------------------
\section{Proposed TCA-Based Trust Management}
We introduce a novel approach that integrates ML with existing trust management frameworks to address the challenges of trust convergence under varying network conditions. Our TCA solution predicts and mitigates the impact of network conditions on trust convergence. This section details our architecture, methodology, and implementation.

%                 ----------------------- Subsection A ----------------------------
\subsection{Solution Architecture}
\begin{figure}[ht]
\vspace{-10pt}
    \centering
    \includegraphics[width=0.96\linewidth]{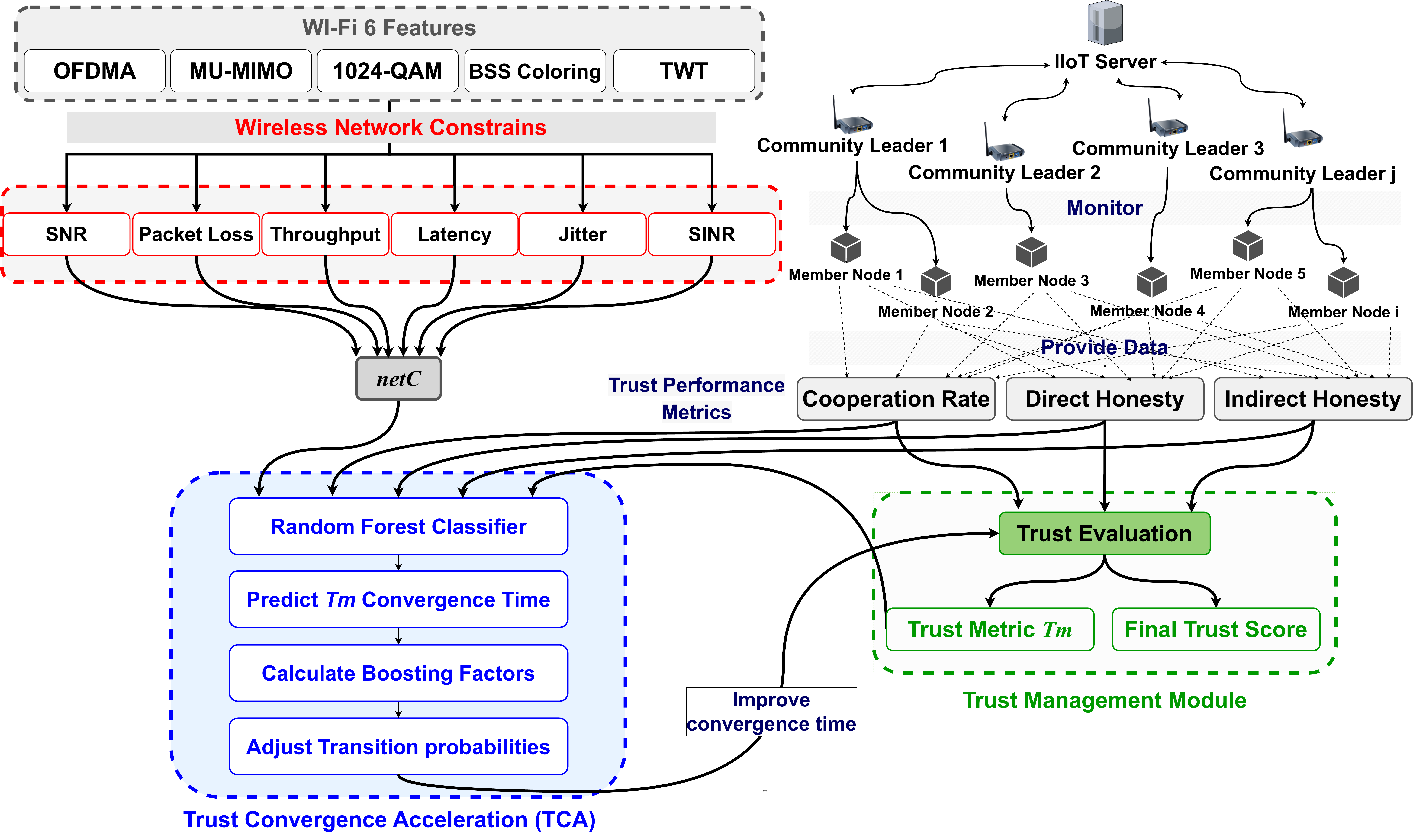}
    \caption{Architecture of our TCA solution.}
    \label{fig:solution_schema}
    \vspace{-10pt}
\end{figure}

Our solution enhances trust management adaptability through the TCA module, which optimizes trust metric convergence through network condition quantification, ML prediction, and dynamic trust adjustment. Fig. \ref{fig:solution_schema} illustrates our approach.

The TCA operates as an enhancement layer that integrates with existing trust frameworks, with three key components:

\begin{enumerate}
    \item \textbf{Network Condition Quantification}: A comprehensive network condition parameter ($netC$) aggregates multiple quality indicators, providing real-time assessment of network health.
    
    \item \textbf{ML Prediction}: A pre-trained Random Forest model predicts convergence time under specific network conditions, enabling proactive trust adjustments.
    
    \item \textbf{Dynamic Trust Adjustment}: Based on predictions, the system adaptively modifies trust evaluation parameters to accelerate convergence while maintaining accuracy.
\end{enumerate}

Importantly, the TCA operates predominantly offline, using pre-trained models to make predictions. This design minimizes computational overhead during operation, making it suitable for resource-constrained IIoT environments.

%                 ----------------------- Subsection B ----------------------------
\vspace{-5pt}
\subsection{Network Condition Parameter ($netC$)}

The $netC$ parameter provides a unified representation of network quality, calculated as a weighted sum of normalized network metrics:

\vspace{-15px}
\begin{equation}
    \label{eq:netC}
    \scalebox{0.9}{$\begin{gathered} netC = \alpha \cdot SNR_{norm} + \beta \cdot (1 - PL_{norm}) + \gamma \cdot (1 - J_{norm}) \\ + \delta \cdot (1 - L_{norm}) + \tau \cdot T_{norm} + \sigma \cdot SINR_{norm}, \end{gathered}$}
\end{equation}

where the normalized metrics represent key QoS parameters, and the weights ($\alpha, \beta, \gamma, \delta, \tau, \sigma$) reflect their relative importance. For this implementation, we assign equal weights ($1/6$ each) to avoid bias toward any single QoS metric without deployment-specific empirical evidence. This baseline approach demonstrated robust performance across all tested industrial scenarios. For metrics where lower values indicate better performance (PL, J, L), we use their complements to maintain consistent polarity in the calculation.
%                 ----------------------- Subsection C ----------------------------
\subsection{Dataset Generation and Preprocessing}
\label{subsec:dataset}

\begin{figure}[ht]
\vspace{-11pt}
\centering
\includegraphics[width=1\linewidth]{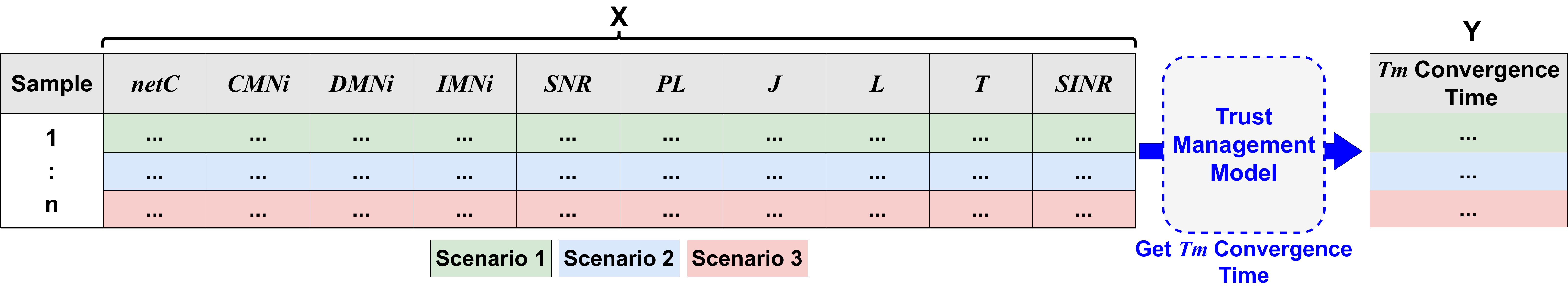}
\caption{Dataset input and output structure.}
\label{fig:dataset_structure}
\vspace{-10pt}
\end{figure}

Building upon the simulation environment introduced in Section \ref{subsec:secIIC}, we utilized MATLAB's WLAN and Communications Toolboxes to generate a comprehensive dataset for ML model training and evaluation. This integration ensured consistency between our network condition modeling and the subsequent ML-based trust acceleration framework, maintaining realistic IEEE 802.11ax channel characteristics throughout both analysis phases. The dataset development process involved the following steps:

\begin{enumerate}
    \item \textbf{Node Selection}: MNs were randomly selected from each industrial community within the simulated environment.
    \item \textbf{Network Condition Simulation}: Each selected node was subjected to varying network conditions, categorized as Good, Medium, and Poor, based on the parameter ranges defined in Table~\ref{tab:simulation_scenarios}.
    \item \textbf{Trust Metric and $netC$ Calculation}: For each node and network condition, we calculated the trust performance metrics and the network condition parameter.
    \item \textbf{Trust Convergence Simulation}: We simulated the trust convergence process for each node under the given network conditions, recording the time units required for the $Tm$ to converge.
    \item \textbf{Data Sample Recording}: Each simulation run generated a data sample comprising the features X = [$netC$, $C_{MN_i}$, $D_{MN_i}$, $I_{MN_i}$, SNR, PL, J, L, T, SINR] and the corresponding output Y (convergence time), as illustrated in Fig.~\ref{fig:dataset_structure}.
\end{enumerate}

The initial dataset comprised approximately 35,000 samples with a wide range of convergence times. To facilitate effective model training and address the inherent variability in trust convergence times, we organized the convergence times into six distinct classes:

\begin{itemize}
    \item Class 1 (Very Fast Convergence): Y = 4
    \item Class 2 (Fast Convergence): Y = 5-6
    \item Class 3 (Moderate Convergence): Y = 7-9
    \item Class 4 (Slow Convergence): Y = 10-12
    \item Class 5 (Very Slow Convergence): Y = 13-15
    \item Class 6 (Extremely Slow Convergence): Y $\geq$ 16
\end{itemize}

We then employed random under-sampling to balance the dataset, resulting in 1,000 samples per class and a total of 6,000 samples. This balanced dataset provides a robust foundation for training and evaluating our ML model.

%                 ----------------------- Subsection D ----------------------------
\subsection{Trust Convergence Acceleration (TCA)}
\label{subsec:tca}

The TCA leverages ML to predict the convergence time of trust metrics based on network conditions and trust parameters. We evaluated several ML models, including Random Forest \cite{Breiman2001}, XGBoost \cite{Chen2016}, Support Vector Machine (SVM) \cite{Cortes1995}, and Logistic Regression \cite{Peng2002}. 
We opted against deep learning models for several reasons specific to IIoT: their substantial computational demands render them impractical for resource-constrained devices; neural networks require significantly larger training datasets than were feasible in our context; deployment across distributed IIoT environments would introduce considerable complexity without commensurate improvements for this prediction task; and traditional ML approaches offer superior interpretability, a crucial factor in security-critical applications.
The dataset was split into 70\% for training, 15\% for validation, and 15\% for testing. Model performance was evaluated using Accuracy, Precision, Recall, and F1-score \cite{chicco2020advantages}, as shown in Table~\ref{tab:ml_comparison}.

\begin{table}[ht]
    \vspace{-5pt}
    \caption{Comparison of ML models.}
    \label{tab:ml_comparison}
    \begin{center}
        \vspace{-5pt}
        \begin{tabular}{|c|c|c|c|c|}
            \hline
            \textbf{Model} & \textbf{Accuracy} & \textbf{Precision} & \textbf{Recall} & \textbf{F1-Score} \\
            \hline
            Random Forest & 92.66\% & 92.72\% & 92.73\% & 92.63\% \\
            \hline
            XGBoost & 91.50\% & 91.65\% & 91.58\% & 91.55\% \\
            \hline
            SVM & 89.12\% & 89.35\% & 89.20\% & 89.18\% \\
            \hline
            Logistic Regression & 85.45\% & 85.60\% & 85.52\% & 85.48\% \\
            \hline
        \end{tabular}
    \end{center}
    \vspace{-10pt}
\end{table}

Based on these results, the Random Forest classifier was selected for the TCA implementation due to its superior performance across all metrics. The Random Forest model achieved the highest accuracy (92.66\%) and F1-score (92.63\%), demonstrating its strong capability in accurately predicting trust convergence time ranges under varying network conditions. Additionally, Random Forest's ensemble approach provides robustness to noise in feature data. The model's inference complexity of $O(T \cdot d_{\max})$ (trees × depth) enables real-time evaluation on CLs with negligible overhead.

%                 ----------------------- Subsection E ----------------------------
\subsection{Dynamic Trust Evaluation Adjustment}

To enhance adaptability to varying network conditions, we introduce a boosting factor ($bf$) that dynamically adjusts trust convergence based on predicted convergence time and current network quality. The boosting factor is computed as:

    \vspace{-5pt}
\begin{equation}
    \scalebox{0.9}{$
    \begin{gathered}
        bf = 1 + (1 - netC) \cdot \lambda \cdot \
        \min\left(\frac{pc}{maxT}, 1\right),
    \end{gathered}$}
\end{equation}

where $pc$ is the predicted class from the ML model, $maxT$ is the maximum convergence time units (for normalization), and $\lambda$ is a scaling factor (empirically set to 0.2 for optimal convergence-stability balance). When network conditions are good ($netC \geq 0.8$), no boosting is applied ($bf = 1$).

Algorithm \ref{alg:boostingFactor} outlines the boosting factor calculation process, which is then used to adjust transition probabilities in the Markov chain trust model, effectively accelerating trust convergence under poor network conditions while maintaining evaluation accuracy.
\begin{algorithm}[ht]
    \fontsize{7}{7}\selectfont
    \caption{Boosting Factor Calculation}
    \label{alg:boostingFactor}
    \KwIn{$netC$, $pc$, $maxT$}
    \KwOut{$bf$}

    % Define class ranges based on predicted class output
    $classRanges \leftarrow [4, 6, 9, 12, 15, maxT]$\tcp*[f]{\textit{Class 1 to 6 upper bounds}}\;
    $pc \leftarrow classRanges[pc]$\tcp*[f]{\textit{Select upper bound of predicted class range}}\;

    \If{$netC \geq 0.8$}{
        $bf \leftarrow 1$\tcp*[f]{\textit{No boosting for good network conditions}}\;
    }
    \Else{
        $baseFactor \leftarrow 1 + (1 - netC) \cdot \lambda$\tcp*[f]{\textit{For $\lambda = 0.2$, the maximum boosting factor is 1.2}}\;
        $classFactor \leftarrow \min\left(\frac{pc}{maxT}, 1\right)$\tcp*[f]{\textit{Normalize predicted class range using upper bound}}\;
        $bf \leftarrow 1 + (baseFactor - 1) \cdot classFactor$\tcp*[f]{\textit{Adjust bf}}\;
    }
    \Return $bf$\;
\end{algorithm}
\setlength{\textfloatsep}{0pt}
This approach allows our system to distinguish between poor network conditions and malicious behavior by adjusting trust evaluation parameters based specifically on network metrics. Since trust calculations incorporate both performance metrics (cooperation, honesty) and network conditions, the system can identify when poor performance is likely due to network issues rather than malicious intent.

%                 ----------------------- Subsection F ----------------------------
\vspace{-4pt}
\subsection{Performance Evaluation}

\begin{figure*}[h]
\centering
\begin{subfigure}[t]{0.29\textwidth}
    \includegraphics[width=\linewidth]{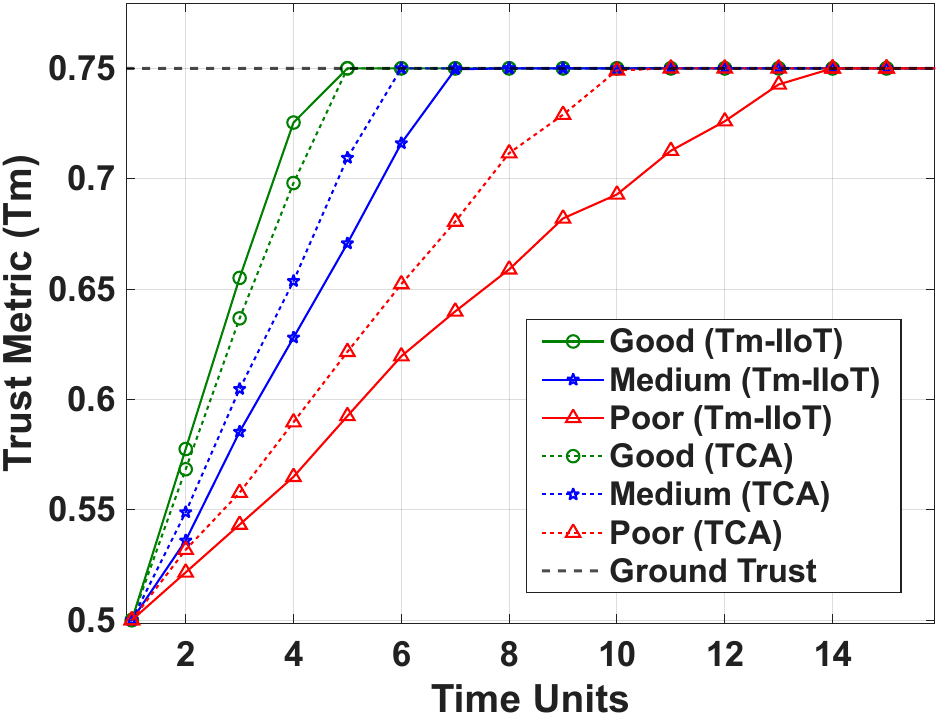}
    \caption{Tm-IIoT vs TCA convergence.}
    \label{fig:convergence_comparison}
\end{subfigure}
\hfill
\begin{subfigure}[t]{0.29\textwidth}
    \includegraphics[width=\linewidth]{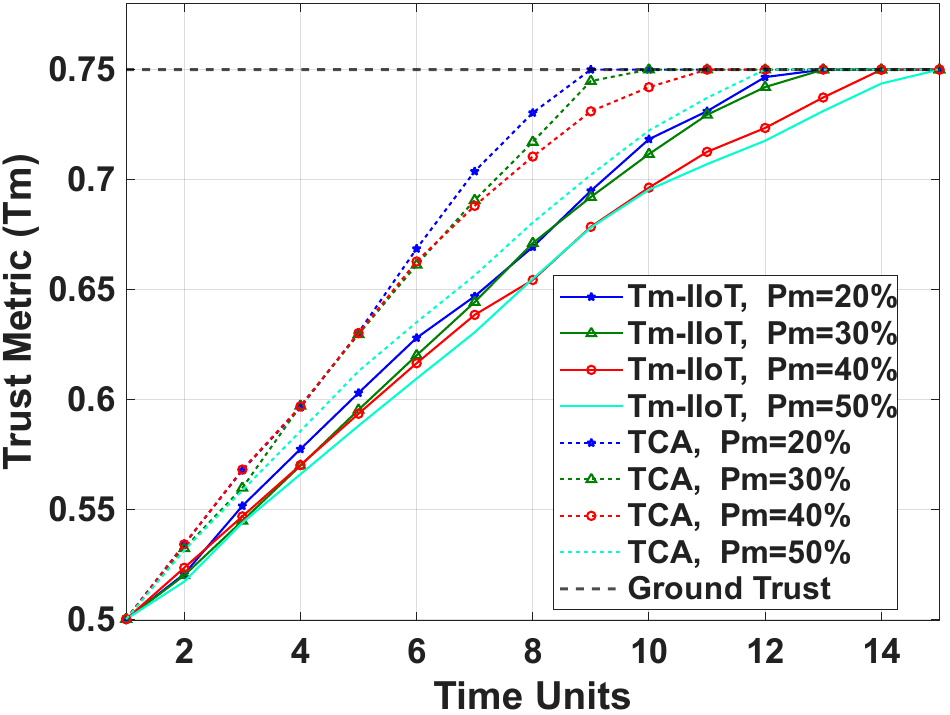}
    \caption{Effect of malicious nodes.}
    \label{fig:convergence_comparison_malicious}
\end{subfigure}
\hfill
\begin{subfigure}[t]{0.277\textwidth}
    \includegraphics[width=\linewidth]{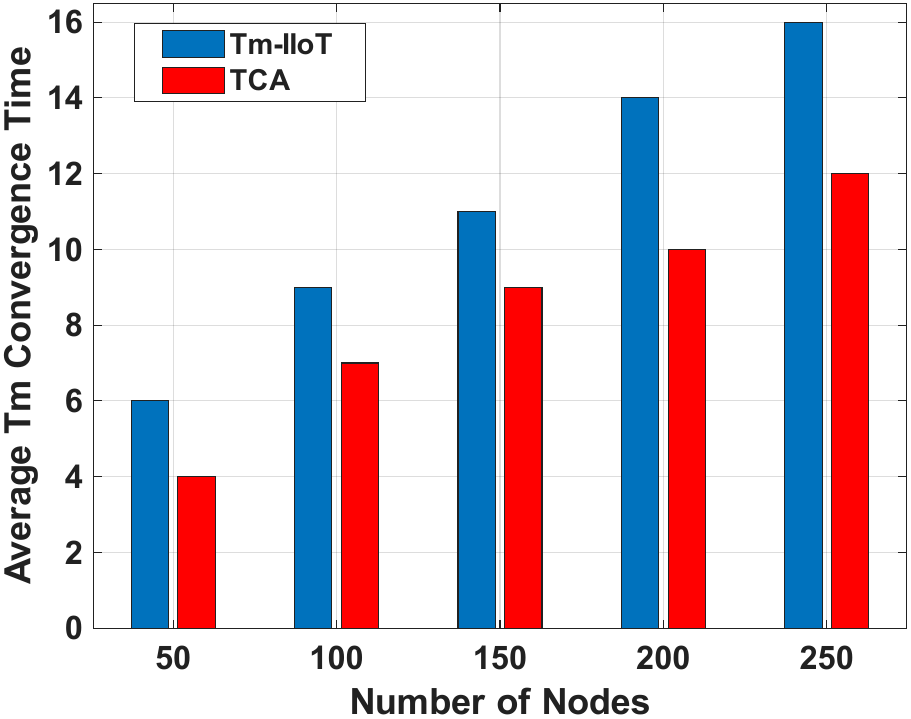}
    \caption{Scalability analysis.}
    \label{fig:scalability_analysis_convergence}
\end{subfigure}
\caption{Performance comparison of the TCA-based trust model.}
\label{fig:performance_figures}
\vspace{-15pt}
\end{figure*}

We conducted comprehensive simulations comparing our TCA-based approach with the original Tm-IIoT model across varying network conditions and in the presence of malicious nodes. The results are presented in Fig. \ref{fig:performance_figures}.
Fig. \ref{fig:convergence_comparison} illustrates trust convergence for a monitored node, comparing TCA with the original Tm-IIoT under different network conditions. In Scenario 1 (Good network conditions), both approaches converge rapidly in approximately 4 time units, as optimal conditions facilitate efficient trust evaluation. However, TCA's advantages become evident in challenging environments. In Scenario 2 (Medium conditions), TCA achieves convergence in about 6 time units versus 7 for Tm-IIoT, yielding a 14.3\% reduction in convergence time. The improvement is most pronounced in Scenario 3 (Poor conditions), where TCA converges significantly faster, requiring approximately 10 time units compared to 14 for Tm-IIoT. This represents a substantial 28.6\% reduction in convergence time, demonstrating TCA's effectiveness in accelerating trust stabilization, particularly when network quality degrades.

To evaluate robustness against malicious behavior, we analyzed performance during bad-mouthing attacks \cite{Buja2023}, where attackers provide false negative feedback to unfairly lower legitimate nodes' trust scores. Fig. \ref{fig:convergence_comparison_malicious} presents comparative results under poor network conditions with malicious node percentages ($Pm$) ranging from 20\% to 50\%. While convergence time increases for both models as attack intensity rises, TCA consistently demonstrates superior performance over Tm-IIoT in terms of convergence speed. At $Pm$=20\% , TCA converges significantly faster, requiring only 9 time units compared to 13 for Tm-IIoT, which represents a 30.77\% reduction. This advantage persists as the percentage of malicious nodes increases: at $Pm$=30\% , TCA converges in 10 time units versus 13 for Tm-IIoT (a 23.08\% improvement), and at $Pm$=40\% , it takes 11 time units for TCA compared to 14 for Tm-IIoT (a 21.43\% reduction). Even under high attack intensity of $Pm$=50\% , TCA maintains a clear edge, converging in 12 time units while Tm-IIoT requires approximately 14.5 time units, achieving a 17.24\% faster convergence. Furthermore, Fig. \ref{fig:convergence_comparison_malicious} confirms the robustness of our approach; while both models eventually converge near the ground truth (0.75), TCA reaches this level significantly faster, demonstrating that the acceleration mechanism does not negatively impact final trust evaluation accuracy, even under coordinated attack scenarios.

Finally, we evaluated scalability by analyzing the average convergence time with increasing network sizes (from 50 to 250 nodes) under poor network conditions, as depicted in Fig. \ref{fig:scalability_analysis_convergence}. The results clearly demonstrate that TCA consistently exhibits lower average convergence times across all tested network scales. For a smaller network of 50 nodes, TCA converges in approximately 4 time units compared to 6 for Tm-IIoT, achieving a significant ~33\% reduction. This performance advantage is maintained as the network grows. For the largest tested network of 250 nodes, TCA requires only 12 time units for convergence versus 16 for Tm-IIoT, still providing a notable 25\% reduction. This confirms that TCA's benefits in accelerating trust convergence are preserved even in large-scale IIoT deployments, highlighting its practical applicability.

\vspace{-5pt}
% --------------------------------------- Section IV ----------------------------------------------
\section{Conclusion and Future Research}
This paper has presented a novel ML-enhanced trust management framework for IIoT environments that effectively addresses the challenges of varying network conditions in industrial settings. Our TCA approach achieves up to 28.6\% faster trust convergence under challenging network conditions while maintaining robust accuracy in detecting malicious nodes. Experimental results demonstrate the framework's resilience against bad-mouthing attacks and excellent scalability across networks of varying sizes, confirming its practical applicability in diverse industrial deployments. While our implementation utilizes IEEE 802.11ax as the foundational standard, the framework's inherent adaptability enables its application across diverse network architectures, including emerging cellular infrastructures. The modular design facilitates deployment as a network layer overlay in IIoT systems, interfacing effectively with standard industrial communication protocols.
Future research will focus on real-world testbed validation using standards-compliant APs as CLs and ESP32-C6 devices as MNs, enhancing threat detection capabilities for sophisticated attack scenarios, and extending support to emerging industrial connectivity paradigms including deterministic networking approaches.
\vspace{-10px}
\section*{Acknowledgment}
This work was funded by the French National Research Agency (ANR-22-PEFT-0007) as part of France 2030 and the NF-FITNESS project.

\vspace{-5px}
\bibliographystyle{ieeetr}
\bibliography{bibfile}
\end{document}